# Dispersion supported BB84 quantum key distribution using phase modulated light

J. Mora, A Ruiz, W. Amaya and J. Capmany

iTEAM Research Institute, Universidad Politécnica de Valencia

Camino de Vera s/n, 46022 – Valencia (Spain)

**ABSTRACT**

We propose and experimentally demonstrate that, contrary to what it was thought up to now, BB84 operation is feasible using the double phase modulator (PM-PM) configuration in frequency coded systems. This is achieved by exploiting the phase to intensity conversion due to the chromatic dispersion provided by the fiber linking Alice and Bob. Thus, we refer to this system as dispersion supported or DS BB84 PM-PM configuration.

Quantum cryptography features the unique way of sharing a random sequence of bits between users with a certifiably security not attainable with either public or secret-key classical cryptographic systems. This is achieved by means of quantum key distribution (QKD) techniques, which rely on exploiting the laws of quantum mechanics [1].

QKD deals with the need to distribute a key between a transmitter (Alice) and a receiver (Bob) with complete confidentiality. If Alice and Bob encode their bits in states of a quantum system, a third party, the eavesdropper (Eve) interested in gaining access to the information they share will modify the quantum state and therefore will destroy its information. At the same time, the eavesdropper will neither be able to keep a perfect copy of the sequence nor to send it again in order to avoid being detected, as this is not allowed by the non-cloning theorem [2].

Photonics is the principal enabling technology for long distance QKD using optical fiber links. A particularly interesting approach is the so called frequency coding technique proposed by Merolla and co-workers [3] which relies on encoding the information bits on the sidebands of either phase [4] or amplitude [5] radiofrequency (RF) modulated light. In essence, Alice randomly changes the phase of the electrical signal used to drive a light modulator among four phase values ($0$, $\pi$) and ($\pi/2$, $3\pi/2$), which form a pair of conjugated basis. When arriving to Bob, he modulates the signal again using the same microwave signal frequency and thus his new sidebands will interfere with those created by Alice [3]. The frequency coded approach has the additional added value that its capacity can be upgraded by adding more microwave subcarriers, an approach which is known as Subcarrier Multiplexed Quantum Key Distribution (SCM-QK) [6]-[8]

The original proposal by Merolla and co-workers [5] based on the use of a pair of simple phase modulators (PM-PM configuration) was, in principle, suited only for the implementation of the B92 protocol [9]. In fact, to demonstrate the implementation of the BB84 protocol it had to be modified by replacing the phase modulators by amplitude modulators (AM-AM configuration) [5,10], which, in addition, need to be properly biased for successful operation.

In this letter we demonstrate, contrary to what it was thought up to now, that BB84 operation is feasible using the PM-PM configuration. This is achieved by exploiting the phase to intensity conversion due to the chromatic dispersion provided by the fiber linking Alice and Bob. Thus, we refer to this system as dispersion supported or DS PM-PM configuration. This result is important for two reasons: first of all because it allows to implement the BB84 protocol rather than the insecure B92 protocol using the simplest frequency coded configuration (PM do not require bias voltage) and secondly because the conditions under which this is possible are compatible with those to be found in practice, since Alice and Bob will be, in general separated by a dispersive optical fiber link.

The system under consideration is shown schematically in Fig. 1. We briefly recall its operation [3]. Assuming that Alice's transmitter contains a monochromatic optical source that emits photons with an angular frequency $\omega_o$. The optical source is externally modulated by a phase modulator (PM1) which is fed by means of a local oscillator OL1.

The local oscillator gives an RF signal of frequency Ω in which Alice can introduce a random phase shift $\Phi_A$ to encode the binary secret key (0 for bit "0" and p for bit "1"). Therefore, the output signal of Alice's transmitter can be described when low modulation index $m_A$ is considered as:

$$E_{ALICE}(t) = E_o e^{-j\omega_o t} \cdot t_A \cdot \left\{ 1 + jm_A \cdot cos(\Omega t + \Phi_A) \right\} \quad (1)$$

Where $E_o$ is the amplitude of the electrical signal related with the average number of photons per bit and $t_A$ represents the optical losses of the PM1. After propagation through a fiber link of length L, the light is again externally modulated by another PM2 at Bob's receiver. PM2 is driven also with a frequency Ω but with variable phase $\Phi_B$ coming from an OL2. Then, the output optical signal after PM2 is given by:

$$E_{BOB}(t) = E_o e^{-j\omega_o t} \cdot e^{j\beta_o L} e^{-\alpha L} t_A t_B \times$$
$$\left\{ 1 + \frac{jm_A}{2} e^{j\frac{1}{2}\beta_2 L \Omega^2} \cdot cos(\Omega t + \Phi_o + \Phi_A) \right\} \left\{ 1 + jm_B \cdot cos(\Omega t + \Phi_B) \right\} \quad (2)$$

where α is the optical fiber losses, $t_B$ is the optical losses of the PM2 and $m_B$ is the modulation index of the PM2. From equation (2), we can observe that fiber chromatic dispersion $\beta_2$ introduces a phase factor $\Phi_o = \beta_2 L$ after propagation between the sidebands and the carrier of the optical signal coming from Alice's transmitter.

The final stage in the system is composed of two optical filters centered at $\omega_o+\Omega$ and $\omega_o-\Omega$ in order to measure the intensity of each optical sideband carrying the secret key.

The output of each filter is then sent to a photon counter. Therefore, the optical power normalized for each sideband is given by:

$$P(\omega_o + \Omega) = \frac{1}{2}\left[1 + V\cos(\Phi_B - \Phi_A)\right]$$
$$P(\omega_o - \Omega) = \frac{1}{2}\left[1 + V\cos(\Phi_B - \Phi_A + \beta_2 L\Omega^2)\right]$$
(3)

Where the parameter V corresponds with visibility of each band, which can be written as:

$$V = \frac{2m_A m_B}{m_A^2 + m_B^2} \tag{4}$$

From equation (3), we observe that both sidebands have the same value when the fiber dispersion is negligible or compensated, i.e., when the term $\beta_2 L\Omega^2$ can be considered close to zero or negligible, as assumed, for instance in [3]. Therefore, the BB84 protocol cannot be implemented since the amplitudes of both sidebands are always equal regardless of which particular base is chosen. However, we can see that the BB84 protocol can be implemented when the visibility is unity ($m_A = m_B$) and the following condition is fulfilled:

$$\beta_2 L\Omega^2 = n\pi \tag{5}$$

Since then, equation (3) can be written as [5]:

$$P(\omega_o + \Omega) = \cos^2\left(\frac{\Phi_B - \Phi_A}{2}\right) \tag{6}$$

$$P(\omega_o - \Omega) = sin^2\left(\frac{\Phi_B - \Phi_A}{2}\right)$$

Equation (5) provides a design criterium for the system. If the overall link dispersion $\beta_2 L$ is fixed, then (5) gives the required value of the subcarrier frequency. On the other hand, if $\Omega$ is fixed, then from (5) we get the value of the minimum required link dispersion.

Implementing the QKD system with a pair of phase modulators brings up several advantages in terms of cost and system complexity. For instance phase modulators are cheaper than amplitude modulators and there is no need to fulfil the counterphase bias condition required when using two amplitude modulators. This point is certainly important since the modulator bias voltage tends to drift with time and keeping this condition would require in practice an additional simultaneous bias tracking circuitry. In addition, we can consider this configuration as a security upgrade since only the BB84 protocol is implemented between Alice and Bob when (5) is fulfilled. For example if $\Omega$ and $\beta_2$ are fixed and Alice and Bob are separated by a distance $L = \pi/\beta_2\Omega^2$ then any attempt of attack made by Eve at a point 0<z<L between Alice and Bob will result in her retrieving information with an increased QBER(z) as compared to that of a standard BB84 protocol. In fact, a simple computation yields the value:

$$QBER(z) = \frac{1}{2}\frac{(1-C)\cdot p^{signal}(z) + d_B}{p^{signal}(z) + d_B} \quad (7)$$

Where $d_B$ is the detector dark count probability and the contrast C is defined as:

$$C(z,\Omega)=\frac{V \cdot cos^2\left(\frac{1}{2}\beta_2 z\Omega^2\right)}{1+V\,sin^2\left(\frac{1}{2}\beta_2 z\Omega^2\right)} \tag{8}$$

Note that the contrast takes its maximum value which corresponds with the visibility V for z=L. As an example, figure 2 shows the evolution of the QBER for the DS PM-PM configuration as a function of the link distance z normalized to the total link length L for the case where $\beta_2 L\Omega^2 = \pi$, $d_B$=8.10$^{-6}$ and V=98%. For the sake of comparison a broken-trace curve is included showing the QBER evolution for the BB84 system based on the AM-AM configuration [merolla] and negligible dispersion. It becomes clear that the dispersion supported scheme converges to the standard BB84 system performance for z=L.

In order to validate the proposal, we have tested the QKD system under classical operation regime (similar results are to be obtained when attenuating the signals as long as they can be represented by coherent states). The experimental demonstration was implemented with an optical laser delivering an output power of 5 dBm at 1550 nm. Both modulators, PM1 and PM2, were electrooptic phase modulators with a 20 GHz electrical bandwidth and a 2.5 dB optical insertion losses. The half-wave voltage was 7.4 V. The modulation index was 0.35 with a RF signal of 15 GHz. The fiber link had a length around 15 km to comply with equation (5) with n=1, and the spectra were recorded using an optical spectrum analyzer (OSA).

Figure 3 shows the power spectrum density measured at the OSA for three different fiber lengths when Bob chooses the correct base for the same state that Alice (left) or the orthogonal state (right) which corresponds with a phase difference $\Phi_B - \Phi_A$ of 0 or π, respectively. In Fig.3 (a), we can observe that BB84 protocol can be implemented when the condition $\beta_2 L \Omega^2 = \pi$ is experimentally satisfied. According to expression (4), the optical sidebands $\omega_o + \Omega$ and $\omega_o - \Omega$ appear or not depending to the constructive o destructive interference imposed by the phases chosen between Alice and Bob. However, the contrast of the interference is reduced when the length of the fiber link is far from that required by Equation (5) as shown in Figs. 3(b) and 3(c) which plot similar results for a fiber lengths of 7.3 km or a few of meters respectively. Due to the optical fiber losses, the optical power of the optical carrier $\omega_o$ is -0.7, -2.6 and -4.0 dBm, respectively.

The theoretical and experimental values of the amplitudes of the detected sidebands are plotted in figures 4.a (upper sideband) and 4.b (lower sideband) respectively as a function of the phase mismatch between Alice and Bob's RF modulating signals. Results are included for the three different fiber link lengths previously considered. An excellent agreement can be observed between theoretical and experimental results. Furthermore, only for the case where the link length is that designed to fulfill with the condition imposed by Eq (5) one can appreciate the complementary characteristic of the sideband amplitudes, which is a distinctive feature of the BB84 operation.

As an additional supporting experimental evidence, figure 5 plots the theoretical and experimental evolution of the contrast function given by (8). Notice again the excellent agreement between both results.

In summary, we have proposed and demonstrated, that by exploiting the phase to intensity conversion, which takes place in an optical fiber link due to chromatic dispersion, BB84 operation is feasible using the PM-PM configuration, which was thought to be valid only up to now for the implementation of the B92 protocol. Experimental results have been provided to support our proposal showing an excellent agreement with the theoretical predictions.

**ACKNOWLEDGMENTS**

The authors wish to thank the Spanish Government support through Quantum Optical Information Technology, a CONSOLIDER-INGENIO 2010 Project, and the Generalitat Valenciana through the PROMETEO 2008/092 Research Excellency Award

# REFERENCES

[1] N. Gisin, G. Ribordy, W. Tittel and H. Zbiden, Rev. Mod. Phys., vol 74, Nº1, p. 145 (2002)

[2] W.K. Wootters and W.H. Zurek, Nature, vol 299, p. 802 (1982)

[3] J-M. Mérolla, Y. Mazurenko, J. P. Goedgebuer, and W. T.Rhodes, Phys.Rev. Lett. 82, p. 1656 (1999).

[4] J-M. Mérolla, Y. Mazurenko, J. P. Goedgebuer, H. Porte and W. T.Rhodes, Opt. Lett. 24, p. 104 (1999).

[5] O. Guerreau, J-M. Mérolla, A. Soujaeff, F. Patois, J. P. Goedgebuer, and F.J. Malassenet, J. of Selected Topycs in Quantum Electron., 9, p. 1533 (2003).

[6] A. Ortigosa-Blanch and J. Capmany, ' Phys. Rev. A, (024305), 2006

[7] J. Capmany, Optics Express, vol. 17, no. 8, pp. 6457–6464, April 2009.

[8] J. Capmany, A. Ortigosa-Balnch, J. Mora, A. Ruiz, W. Amaya and A. Martínez, J. of Selected Topycs in Quantum Electron., 15, p. 1607 (2009).

[9] C.H. Bennett., Phys. Rev. Lett. 68, p. 3121 (1992).

[10] J-M. Mérolla, L. Duraffourg, J. P. Goedgebuer, A. Soujaeff, F. Patois and W. T.Rhodes, Eur. Phys. J. D 18, p. 141 (2002).

**FIGURE CAPTIONS**

Figure 1. Scheme of the frequency coded system using a pair of phase modulators (PM-PM configuration)). Alice an Bob are separated by a dispersive link of length L.

Figure 2. (Solid trace) Evolution of the QBER for the DS PM-PM configuration as a function of the link distance z normalized to the total link length L for $\beta_2 L \Omega^2 = \pi$ and V=98%. (Broken-trace) QBER evolution for the BB84 system based on the AM-AM configuration [merolla] and negligible dispersion

Figure 3. Power spectra for a fiber link with a length (a) 15 km, (b) 7.3 km and (c) 0 km . Left column results are for $\Phi_A - \Phi_B = 0$. Right column results are for $\Phi_A - \Phi_B = \pi$

Figure 4. Amplitude of sidebands for different phase shifts between Alice and Bob after 15 km (■), 7.3 km (●) y 0 km (▲): (a) left optical sideband and (b) right optical sideband. (Solid traces represent theoretical results) .

Figure 5. Theoretical and experimental evolution of the contrast function given by (8)

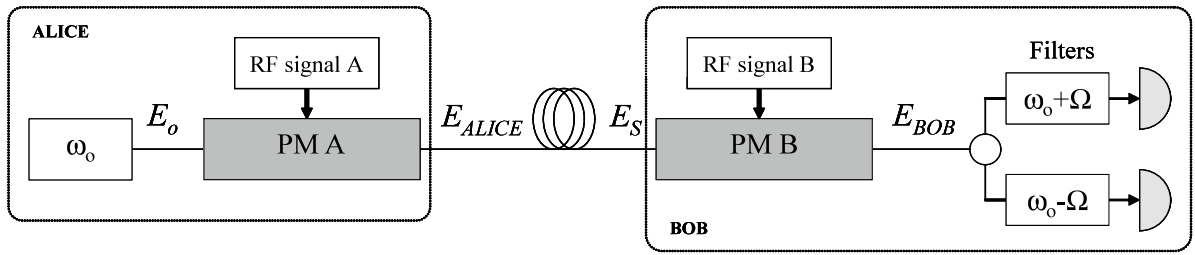

Figure 1. Scheme of the frequency coded system using a pair of phase modulators (PM-PM configuration)). Alice an Bob are separated by a dispersive link of length L

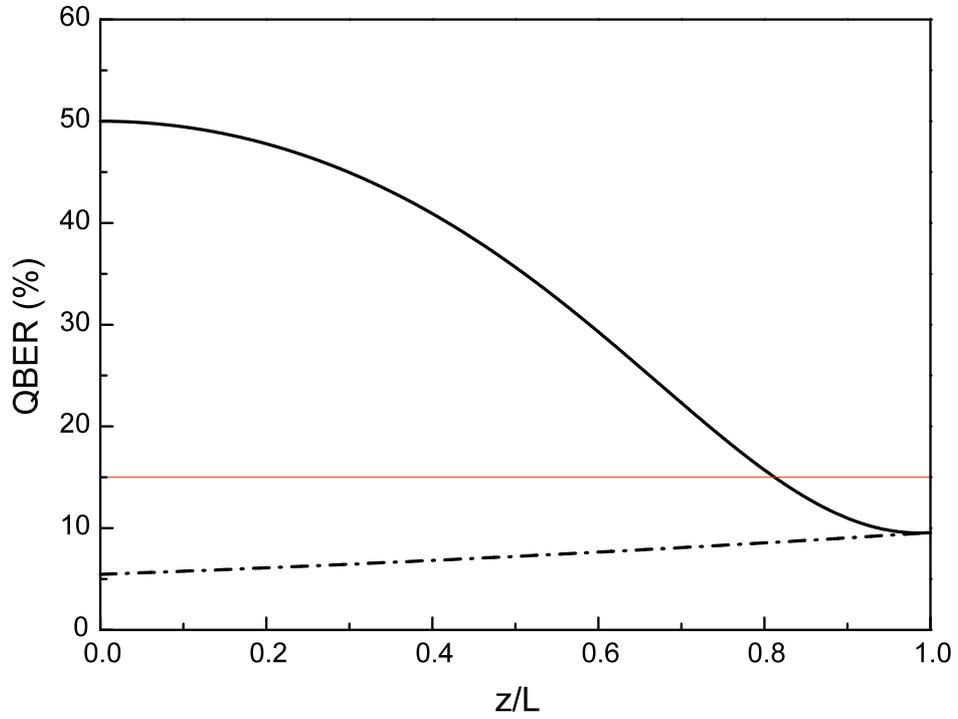

Figure 2. (Solid trace) Evolution of the QBER for the DS PM-PM configuration as a function of the link distance z normalized to the total link length L for $\beta_2 L\Omega^2 = \pi$ and V=98%. (Broken-trace) QBER evolution for the BB84 system based on the AM-AM configuration [5] and negligible dispersion

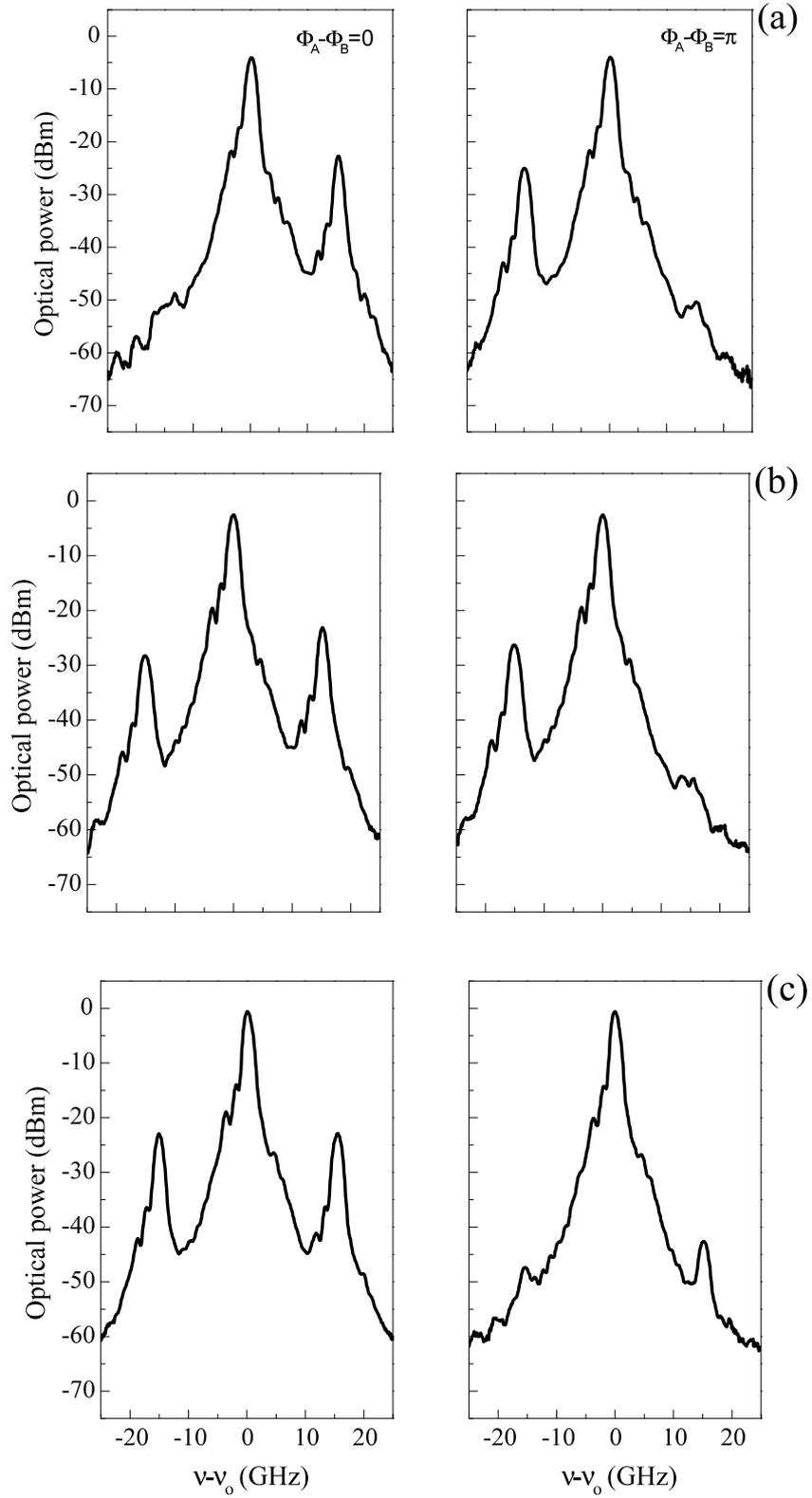

Figure 3. Power spectra for a fiber link with a length (a) 15 km, (b) 7.3 km and (c) 0 km . Left column results are for $\Phi_A - \Phi_B = 0$ . Right column results are for $\Phi_A - \Phi_B = \pi$

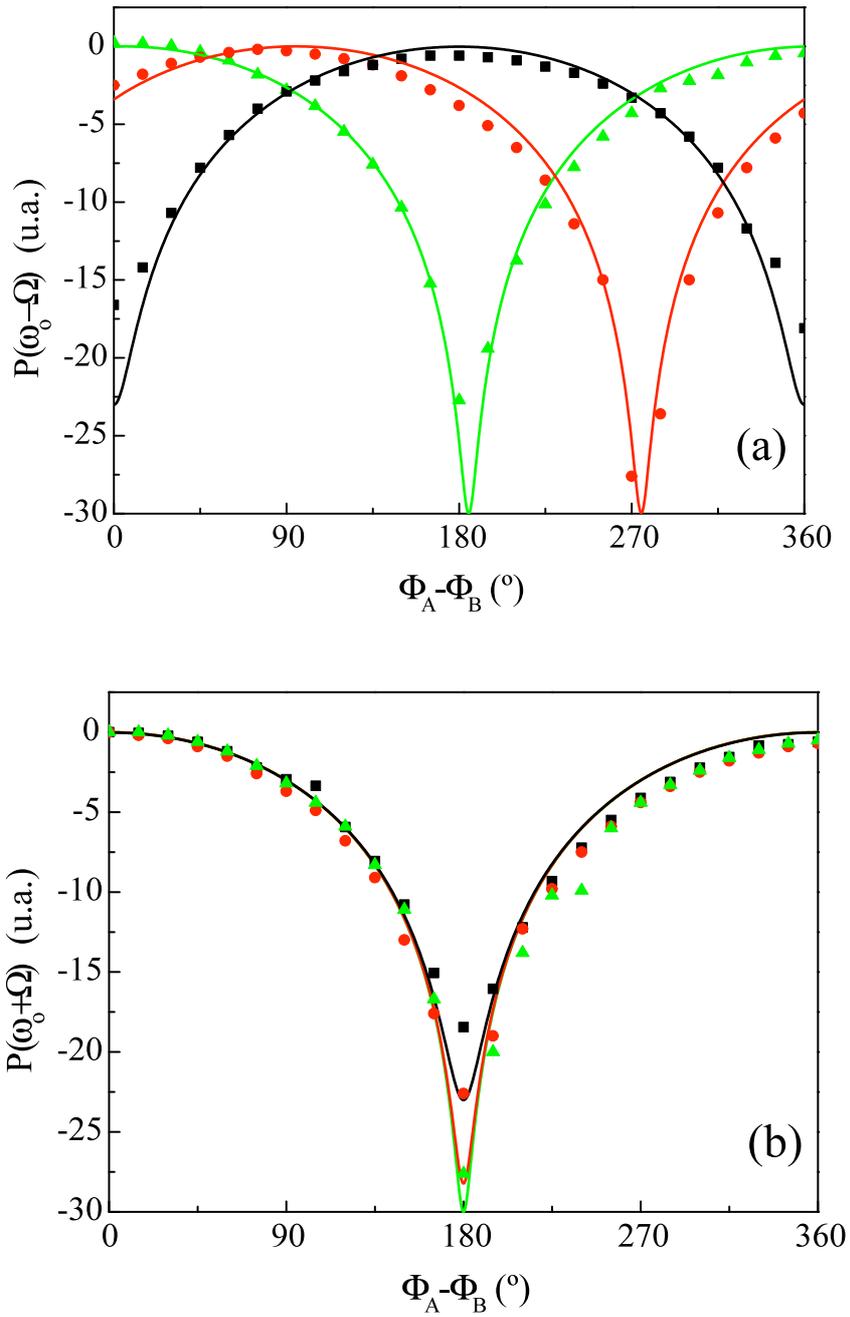

Figure 4. Amplitude of sidebands for different phase shifts between Alice and Bob after 15 km (■), 7.3 km (●) y 0 km (▲): (a) left optical sideband and (b) right optical sideband. (Solid traces represent theoretical results)

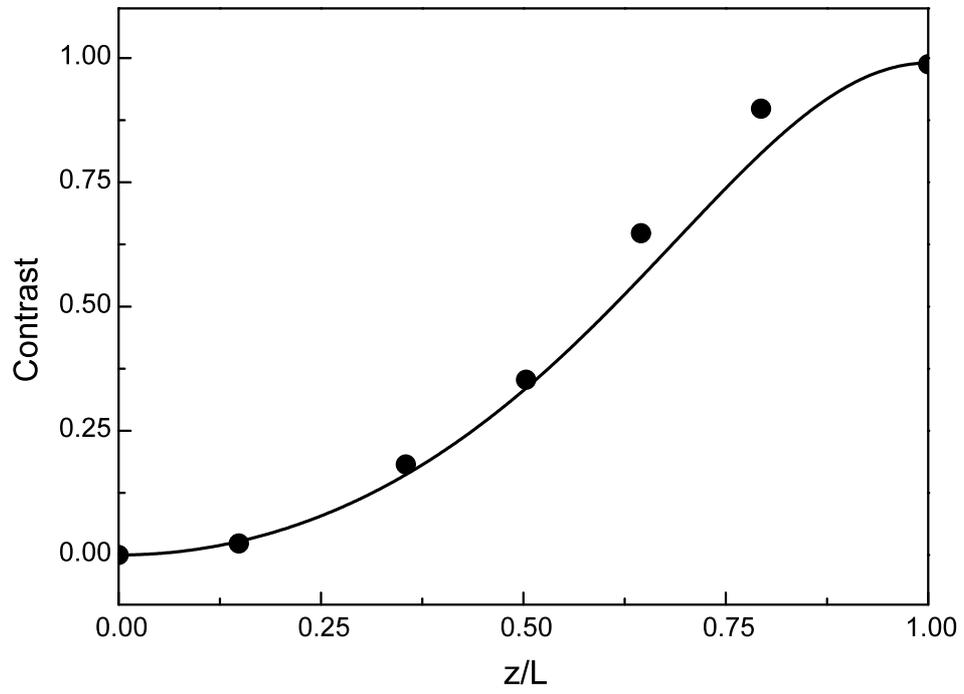

Figure 5. Theoretical and experimental evolution of the contrast function given by (8)